# Orthorhombic metal carbide-borides MeC$_2$B$_{12}$ (Me=Mg, Ca, Sr) from first principles: structure, stability and mechanical properties

Oleksiy Bystrenko [a,1], Jingxian Zhang* [a,b], Tianxing Sun[a,c], Hu Ruan[a,b], Yusen Duan[a,b], Kaiqing Zhang[a], Xiaoguang Li[a]

[a]State Key Laboratory of High Performance Ceramics and Superfine Microstructures, Shanghai Institute of Ceramics, Chinese Academy of Sciences, Shanghai 200050, China;

[b]Center of Materials Science and Optoelectronics Engineering, University of Chinese Academy of Sciences, Beijing 100049, China

[c]University of Chinese Academy of Sciences, Beijing 100049, China;

Abstract:

First principle DFT simulations are employed to study structural and mechanical properties of orthorhombic B$_{12}$-based metal carbide-borides. The simulations predict the existence of Ca- and Sr- based phases with the structure similar to that of experimentally observed earlier compound MgC$_2$B$_{12}$. Dynamical stability of both phases is demonstrated, and the phase CaC$_2$B$_{12}$ is found to be thermodynamically stable. According to simulations, Ca- and Sr- based phases have significantly enhanced mechanical characteristics, which suggest their potential application as superhard materials. Calculated shear and Young's moduli of these phases are nearly 250 and 540 GPa, respectively, and estimated Vickers hardness is 45-55 GPa.

## I. Introduction

Orthorhombic B$_{12}$-based metal borides and related structures attracted remarkable attention of researchers in recent decades. The typical representative of this family, AlMgB$_{14}$ (Imma [74]) (BAM), has the considerable hardness of nearly 30 GPa and low density of 2.7 g/cm$^3$, which, along with low fabrication cost, makes it promising material for various industrial applications [1-3].

In order to explain high hardness of orthorhombic metal borides and to find ways to improve their mechanical properties, a number of theoretical works has been done, in particular, aimed at elucidating the effects of replacing the elements on metal cites and modifying boron network [4-6]. The conclusion important for understanding hardness of BAM-related structures has been formulated in Ref. [7], where, on the basis of the calculations of cleavage strength, it was found, that their shear strength and hardness is limited by relatively weak overall bonding between the layers forming the boron network. Later this conclusion has been illustrated by direct computer simulations in Ref. [8].

[1] Permanent address: Frantsevich Institute for Problems in Materials Science, National Academy of Sciences of Ukraine, Kiev 03142, Ukraine
* Corresponding author: jxzhang@mail.sic.ac.cn

In this study we will focus predominantly on the structural and mechanical properties of closely related compounds, orthorhombic metal carbide-borides of the composition $MeC_2B_{12}$ with taking as metal atoms the elements of II group Mg, Ca, Sr. They can be viewed as the result of the modification of BAM structure by removing metal atoms on Al-sites and replacing inter-icosahedral boron atoms by carbon atoms. In the above context, such a change may give rise to an enhancement of mechanical properties, in particular, due to reducing the distance between boron layers.

The compound $MgC_2B_{12}$ of orthorhombic geometry (Imma [74]) has been earlier synthesized and examined both experimentally and theoretically [4, 9, 10] and was found to be thermodynamically stable. Theoretical prediction for its Young's modulus obtained by first principle simulations (490-500 GPa) is significantly higher than that of BAM compound (430-440 GPa), which suggests high intrinsic hardness of that phase.

At the same time, as far as is known to the authors, Ca- and Sr- based carbide-borides with similar structure have yet not been considered in literature. The purpose of this work is to theoretically study these phases, their stability, structural and mechanical properties in comparison to basic reference phase $MgC_2B_{12}$. We address these issues by means of first principle computer simulations based on density functional theory (DFT).

## II. Computer simulations

Initial data about the structure of $MgC_2B_{12}$ compound (Imma [74]) needed for study was taken from the materials database [11]. This structure is represented by boron layers formed by boron icosahedra $B_{12}$, inter-icosahedral carbon atoms connecting the icosahedra within a layer, and metal atoms embedded on metal sites. In simulations, we used both conventional and primitive unit cells containing 60 and 30 atoms, respectively (Fig.1); both approaches yielded very close results.

First principle simulations were based on density functional theory (DFT) approach implemented in free Quantum Espresso software package [12]. Calculations were performed with the use of pseudopotentials taken from the standard library SSSP, precision version 1.1.2 [13], which employs the generalized gradient approximation (GGA) and exchange-correlation interaction in Perdew–Burke-Ernzerhof (PBE) form [14]. Sampling of the Brillouine zone was done within the framework of Monkhorst-Pack approach [15], predominantly on uniform 5x3x3 and 5x5x5 **k**-point grids in reciprocal space for conventional and primitive unit cells, respectively. The energy cutoff was set 55 Ry for plane wave basis functions, and 450 Ry for charge; the energy convergence threshold used for the electron wave functions was $10^{-6}$ Ry. In addition, to check the accuracy of calculations, selected runs were performed on 6x6x6 and 4x4x4 **k**-point grids, for the energy convergence threshold $10^{-7}$, and cutoff radii 75 Ry and 650 Ry for energy and charge, respectively.

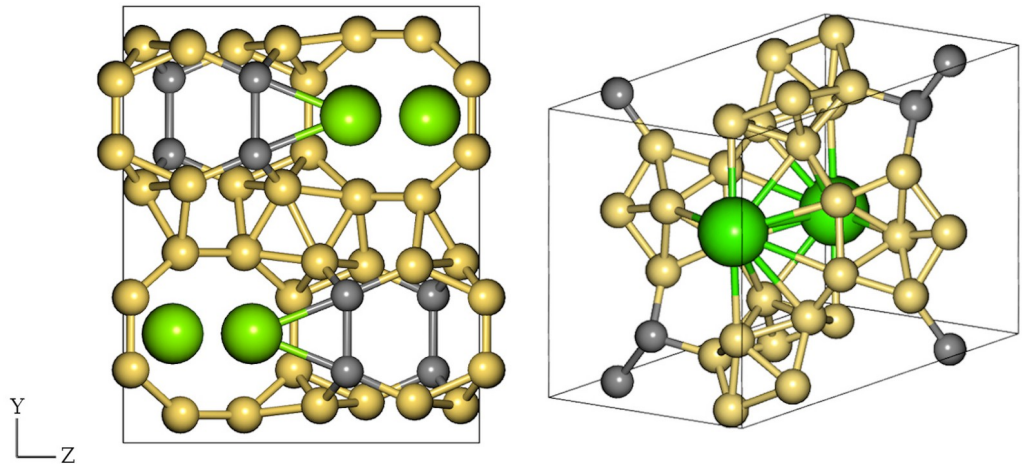

Fig. 1. Conventional (left) and primitive (right) unit cells for $MgC_2B_{12}$ with 60 and 30 atoms, respectively, used in simulation. Magnesium, carbon and icosahedral boron atoms are given in green, gray and yellow, respectively.

Initial configurations for Ca- or Sr-containing phases were prepared from the known $MgC_2B_{12}$ structure by replacing magnesium by Ca or Sr atoms. After that, in order to obtain the equilibrium configurations associated with the minimum energy at zero temperature and pressure, the relaxation procedure by using BFGS algorithm [16] was applied. The relaxation was performed with respect to all ionic coordinates and all lattice parameters (lattice dimensions and angles). The phases obtained in such a way were then examined in detail for thermodynamical and mechanical stability, structural, mechanical and electronic properties.

The structural parameters along with some other quantities obtained in simulations are given in Table I, and the calculated X-ray diffraction (XRD) patterns are displayed in Fig.2.

The crystal system, symmetry and other structural properties of the phases obtained in simulations were determined by free software THERMO_PW [17] and FINDSYM [18] with the tolerance of 0.001 angstrom for ionic coordinates; the formation energy was evaluated with respect to pure elemental phases of carbon as α-graphite ($P6_3/mmc$), boron as α-$B_{12}$ (R-3m), hexagonal magnesium ($P6_3/mmc$), cubic α-calcium (Fm-3m), and cubic α-strontium (Fm-3m).

| No | Compo-sition | Crystal system, symmetry, space group number | Supercell parameters, angstrom | Supercell volume; (vol. per atom), ang$^3$ | Density g/cm$^3$ | Cohesive energy per atom eV | Formation energy per atom eV |
|---|---|---|---|---|---|---|---|
| 1 | MgC$_2$B$_{12}$ | Ortho-rhombic Imma [74] | a=5.61 b=9.81 c=7.93 | 436.29 (7.27) | 2.71 | -6.725 | -0.168 |
| 2 | CaC$_2$B$_{12}$ | Ortho-rhombic Imma [74] | a=5.61 b=10.02 c=8.04 | 451.66 (7.53) | 2.85 | -6.834 | -0.250 |
| 3 | SrC$_2$B$_{12}$ | Ortho-rhombic Imma [74] | a=5.63 b=10.20 c=8.15 | 467.58 (7.79) | 3.43 | -6.744 | -0.180 |

Table I. Structural and other general properties of MgC$_2$B$_{12}$ and related metal carbide-boride phases, obtained in DFT simulations.

As is seen from Table I, the Ca- and Sr-based structures retain the crystal structure of the compound MgC$_2$B$_{12}$ after relaxation, namely, the orthorhombic crystal system and space symmetry Imma. Their formation energies with respect to pure elements are negative, which suggests that these phases may be thermodynamically stable. In order to examine this issue in more detail, we calculated energy for all possible routes of decay to elemental phases specified above, along with known stable phases of CaB$_4$ (P4/mbm), CaB$_6$ (Pm-3m), SrB$_6$ (Pm-3m), SrC$_6$ (P6$_3$/mmc) and two boron carbide phases B$_4$C (R-3m) and B$_{13}$C$_2$ (R-3m).

The compound CaC$_2$B$_{12}$ in all cases turned out to be thermodynamically stable, whereas for the phase SrC$_2$B$_{12}$ the reactions of decay with positive energy were detected, for instance,

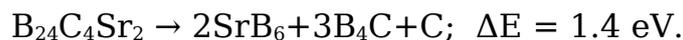

$$B_{24}C_4Sr_2 \rightarrow 2SrB_6 + 3B_4C + C; \quad \Delta E = 1.4 \text{ eV}.$$

This means that this latter phase is thermodynamically unstable and can potentially be only in a metastable state at low temperature and zero pressure. This conclusion correlates with the cohesive and formation energies given in Table I, since the larger numbers have been obtained for the stable phase CaC$_2$B$_{12}$.

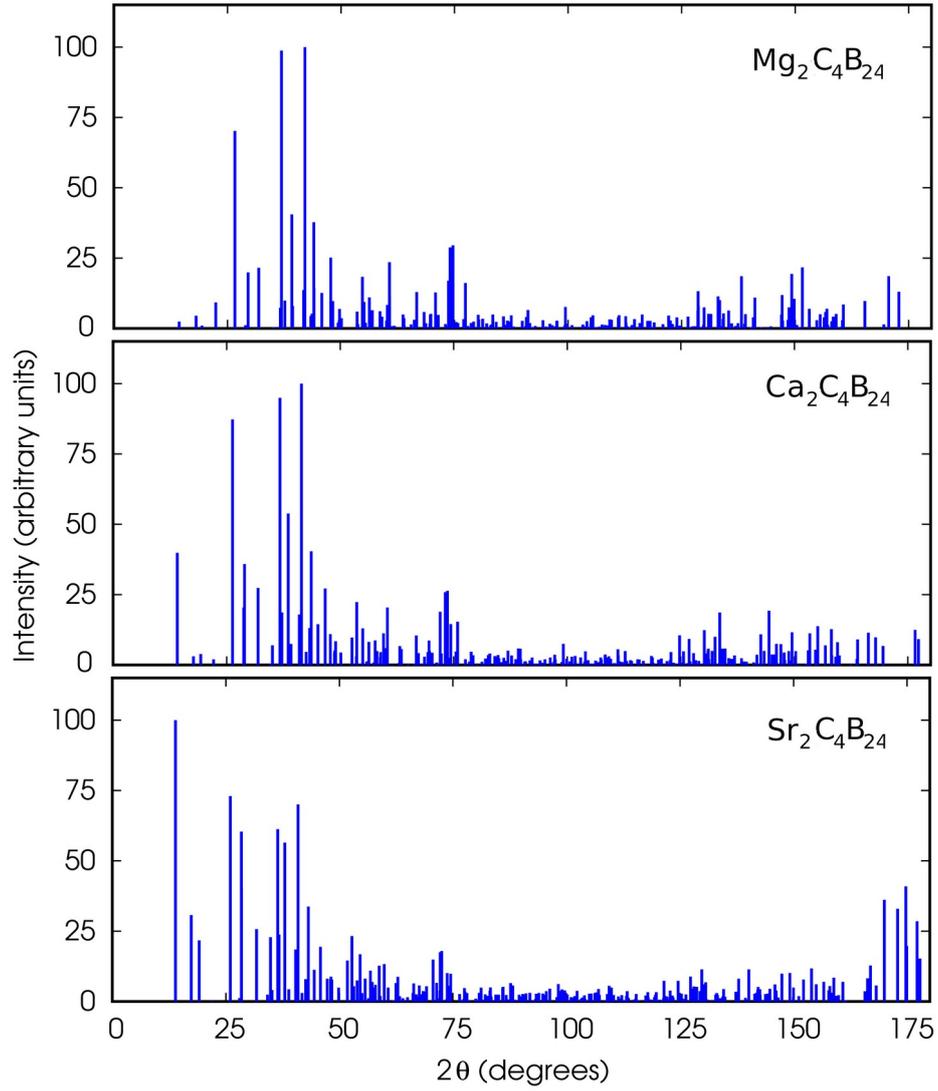

Fig.2. Calculated XRD spectra for $MgC_2B_{12}$ and related compounds.

| No | Compo-sition | Elastic constants, GPa | | | | | | | | |
|---|---|---|---|---|---|---|---|---|---|---|
| | | $c_{11}$ | $c_{22}$ | $c_{33}$ | $c_{44}$ | $c_{55}$ | $c_{66}$ | $c_{12}$ | $c_{13}$ | $c_{23}$ |
| 1 | $MgC_2B_{12}$ | 535 | 560 | 567 | 172 | 278 | 180 | 35.9 | 110 | 57.5 |
| 2 | $CaC_2B_{12}$ | 526 | 548 | 618 | 239 | 276 | 218 | 50.5 | 62.1 | 52.4 |
| 3 | $SrC_2B_{12}$ | 517 | 495 | 627 | 260 | 262 | 231 | 53.9 | 41.7 | 58.0 |

Table II. Elastic (stiffness) constants for $MgC_2B_{12}$ and related compounds, obtained in computer simulations.

| No | Compo-sition | Elastic moduli, GPa | | | Poisson's ratio, $\nu$ | Debye temperature, K | Vickers hardness GPa |
|---|---|---|---|---|---|---|---|
| | | Bulk modulus B | Shear modulus G | Young's modulus, E | | | |
| 1 | $MgC_2B_{12}$ | 231.2 | 218.6 | 498.6 | 0.14 | 1507 | 37.9 (40.8) |
| 2 | $CaC_2B_{12}$ | 224.6 | 247.2 | 542.6 | 0.10 | 1544 | 45.9 (53.0) |
| 3 | $SrC_2B_{12}$ | 214.7 | 248.7 | 538.3 | 0.08 | 1394 | 47.1 (56.9) |

Table III. Mechanical properties of $MgC_2B_{12}$-related compounds, obtained in computer simulations. The numbers given for elastic moduli are Voigt-Reuss-Hill averages; theoretical Vickers hardness was evaluated from elastic constants on the basis of the models given in Ref.[19] and Ref. [20] (in round brackets).

Elastic constants, elastic moduli and Poisson's ratio calculated on the basis of DFT simulations and predicted Vickers hardness are given in Tables II and III. Elastic properties were determined with the use of THERMO_PW [17] software with employing for this purpose the conventional strain-stress approach. Theoretical estimates for Vickers hardness were evaluated from elastic constants on the basis of the empirical models proposed in Refs. [19] and [20]. It is worth mentioning that at present a large number of approaches is developed aimed at predicting hardness from the data obtained from first principle calculations (see, for instance, [21-23]). However, hardness, as a macroscopically defined quantity, may considerably depend on a number of conditions like the way of sample preparation, specificity of indentation, presence of defects, porosity, microstructure, etc. As a result, such models mostly remain semi-empirical, since require additional assumptions or parameters, and, at the same time, their overall predictive accuracy remains on the same order as that of simple above cited models, which we used in this work.

Most interesting finding is that the shear and Young's moduli of the Ca- and Sr-based phases demonstrate significant increase, and the Poisson's ratio, respectively, decrease, as compared to the basic reference Mg-based phase. It is to mention that such a correlation of elastic characteristics is typical for ultrahard materials (like gamma-boron or diamond), and the theoretical estimates given in Table (45-55 GPa) illustrate this dependence.

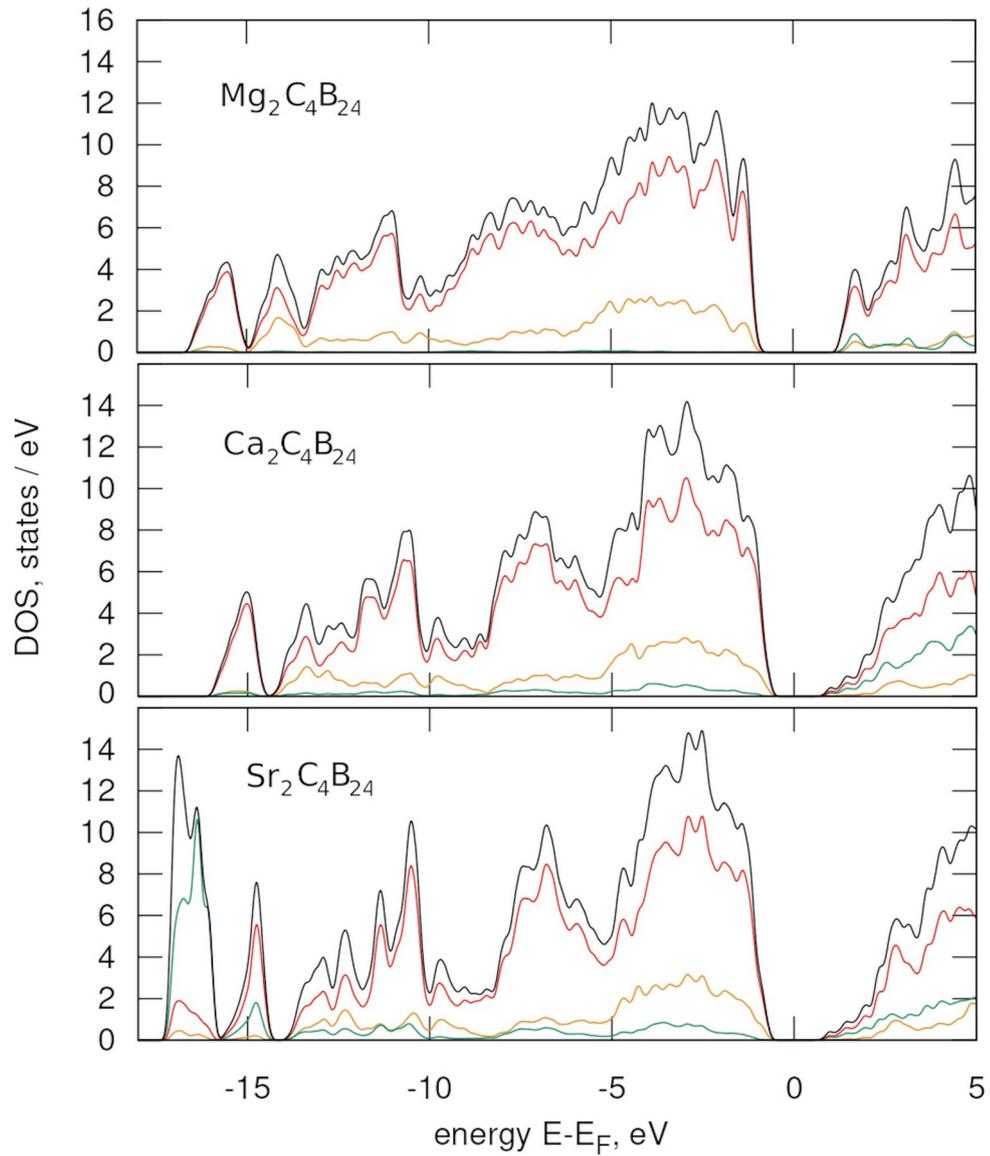

Fig.3. Electronic density of states for $MgC_2B_{12}$-related compounds obtained in simulations. Total DOS is given in black; the curves in red, yellow, and green relate to partial DOS for B, C and metal element, respectively.

Comparison of calculated electronic density of states (DOS) for $MgC_2B_{12}$-related compounds is given in Figs 3. As can be seen from this figure, all $MgC_2B_{12}$-related phases under consideration demonstrate the band gaps of about 1.5-2 eV near the Fermi level typical for semiconductors. Notice that the Mg-related partial DOS (PDOS) within the valence band is nearly zero, which is consistent with purely ionic nature of magnesium bonding in $MgC_2B_{12}$ [4]. At the same time, Ca and Sr demonstrate marked contributions to DOS in valence bands in respective phases reflecting the formation of covalent bonds. The enhancement of predicted mechanical properties of these phases may be due to this fact.

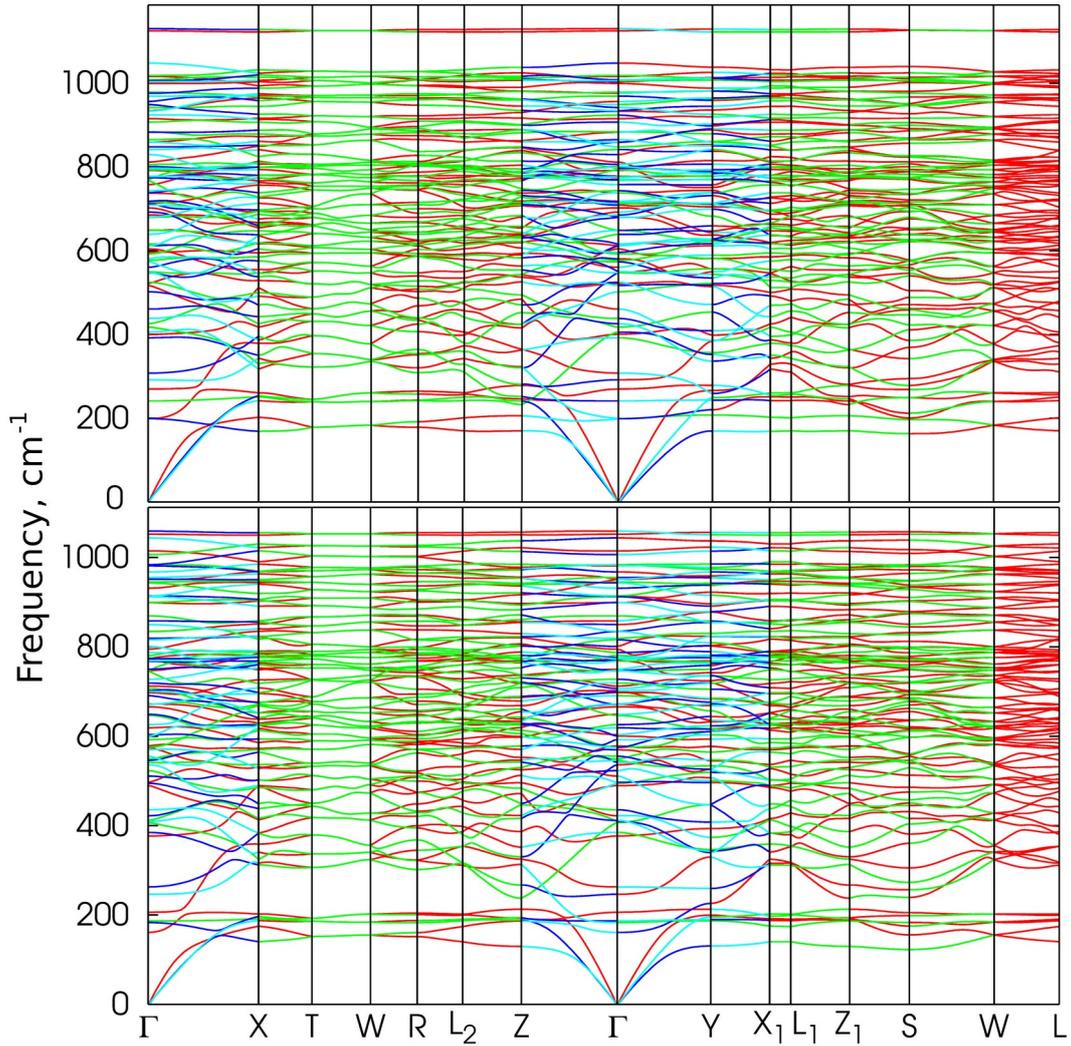

Fig. 4. Calculated phonon dispersion curves for $CaC_2B_{12}$ (top) and $SrC_2B_{12}$ (bottom).

One more important issue concerning theoretically predicted phases is that of their mechanical stability. Static mechanical stability (with respect to static strains) can be easily verified by employing calculated elastic constants to satisfy stability condition for orthorhombic systems, which is a generalization of Born stability criterion for cubic crystals [24]. The dynamical stability, which reflects the property of a crystal structure to remain stable with excitation of individual phonon modes, can be established by the behavior of phonon dispersion curves. We calculated the phonon dispersion curves for Ca- and Sr-based phases, and the results are given in Fig. 4. The absence of unstable modes with negative (imaginary) frequencies indicate the dynamical mechanical stability of both phases.

## Conclusions

To conclude, the first principle DFT computer simulations strongly indicate the existence of Ca- and Sr- based orthorhombic phases, with space group Imma [74], with the structure similar to the experimentally observed reference compound $MgC_2B_{12}$. According to the results of simulations, all $MgC_2B_{12}$-related phases are semiconductors with the band gaps of about 1.5-2 eV. The phase $CaC_2B_{12}$ is found to be thermodynamically stable, which implies that there should be no difficulties in its synthesis. At the same time, the phase $SrC_2B_{12}$ is likely to be metastable at low temperature and zero pressure. Simulations predict mechanical static and dynamical stability of both Ca- and Sr- based phases and significantly enhanced mechanical characteristics as compared to the reference phase, which give the reason to consider them as potentially superhard materials. Evaluated in calculations shear and Young's moduli are nearly 250 and 540 GPa, and the Poisson's ratio and hardness are within the range 0.08-0.1 and 45-55 GPa, respectively.

## Acknowledgment

The authors gratefully acknowledge the support of this research within the framework of the CAS President's International Fellowship Initiative, grant No 2020VEB0005.